\documentclass[twocolumn,           
               showpacs,            
               preprintnumbers,     
               aps,                 
               prd,          	    
               letterpaper,             
               groupaddress,      
               nofootinbib,         
               tightenlines,        
               floats,floatfix      
               ]{revtex4-1}

\usepackage{graphicx}
\usepackage{latexsym}
\usepackage{epstopdf}
\usepackage{amsmath,amssymb}        
\usepackage[draft=false]{hyperref}

\begin{document}

\title{Black holes and the absorption rate of cosmological scalar fields}%

\author{L. Arturo Ure\~na-L\'opez}%
\email{lurena@fisica.ugto.mx}
\affiliation{Departamento de F\'isica, DCI, Campus Le\'on, Universidad
    de Guanajuato, 37150, Le\'on, Guanajuato, M\'exico}

\author{Lizbeth M. Fern\'andez}%
\email{lizbeth@fisica.ugto.mx}
\affiliation{Departamento de F\'isica, DCI, Campus Le\'on, Universidad
    de Guanajuato, 37150, Le\'on, Guanajuato, M\'exico}
\date{\today}%

\pacs{04.40.-b,04.25.D-,95.35.+d,95.36.+x}
\keywords{Scalar fields, black holes}

\begin{abstract}
We study the absorption of a massless scalar field by a static black
hole. Using the continuity equation that arises from the Klein-Gordon
equation, it is possible to define a normalized absorption rate
$\Gamma(t)$ for the scalar field as it falls into the black hole. It
is found that the absorption mainly depends upon the characteristics
wavelengths involved in the physical system: the mean wavenumber and
the width of the wave packet, but that it is insensitive to the scalar
field's strength. By taking a limiting procedure, we determine the
minimum absorption fraction of the scalar field's mass by the black
hole, which is around $50\%$.
\end{abstract}

\maketitle


\section{\label{sec:introduction}Introduction}
Black holes, a concept that emerged from the simplest exact solution
of Einstein's equations, are some of the most fascinating objects in
gravitational physics. Equally fascinating is our current belief that
most galaxies must host a \emph{supermassive} black hole (SMBH) in
their center, with mass values in the range of
$10^5$ to $10^{10}$ solar masses, most likely in a state of very low
matter accretion
nowadays\cite{Caramete:2011eh,*Volonteri:2011rm,*Ferrarese:2002ct,*Nowak:2007ba,*Greenwood:2005cs,Ghez:2000ay}. In
particular, the measurements of the velocities of stars near the
center of the Milky Way have provided strong evidence for the presence
of a SBH with a mass of around $4\times 10^6 M_\odot$\cite{Ghez:2000ay}.

There are some models that attempt to explain the present existence of
galactic SMBH's. Among others, we can mention the collision of two or
more black holes to form a larger one, the core-collapse of a stellar
cluster, and the formation of primordial black holes directly out from
the primordial plasma in the first instants of time after the Big
Bang\cite{Melia:2007vt,*Carr:2009jm}.

The key point in the discussion are the features of the precise
mechanism under which a black hole can accrete enough matter to become
supermassive. In particular, some authors have proposed that
primordial black holes (PBH) can go supermassive simply by accreting
matter from a cosmological scalar field related to dark energy
(quintessence). In a first study, the authors in\cite{Bean:2002kx}
(see also\cite{Babichev:2004yx,*Babichev:2005py,MersiniHoughton:2008aw,Rodrigues:2009eg})
found that PBH could have effectively accreted enough matter from a
quintessence field endowed with an exponential potential.

The calculations for the accretion were in fact based upon the simple
and exact results of the accretion of a massless scalar field into a
black hole found in\cite{Jacobson:1999vr}, see
also\cite{Frolov:2002va,Harada:2004pf,Rodrigues:2009eg}. However, the
results in\cite{Bean:2002kx} were later refused
in\cite{Custodio:2005en}, where was shown that the quintessence flux
must decrease slower than $t^{-2}$ for PBHs to grow at all. This same
result seems to have been confirmed by other authors under more
general assumptions\cite{Frolov:2002va,Harada:2004pf,Carr:2010wk}.

On the other hand, a related topic is the use of a cosmological scalar
field as model for dark matter in the
Universe\cite{Sahni:1999qe,*Matos:2000ng,*Matos:2000ss,*Arbey:2001qi,*Matos:2008ag,*Arbey:2006it,*Liddle:2006qz},
and the possibility that they can be the dominant matter in galaxy
halos\cite{Sin:1992bg,*Ji:1994xh,*Arbey:2003sj,*Alcubierre:2003sx,*Guzman:2003kt,*Matos:2007zza,*Bernal:2009zy,*Barranco:2010ib,*UrenaLopez:2010ur}. If
so, then one has to address the accretion of this dark matter scalar
field into the central SBH that seems to be present in most
galaxies\cite{UrenaLopez:2002du,CruzOsorio:2010qs}.

The aim of this paper is to present some simple results of the
interaction of a scalar field with a black hole, with numerical
calculations based upon previous works in the
literature\cite{Marsa:1996fa,*Thornburg:1998cx,*Choptuik:2003} that
may be useful in the understanding of the accretion, in general terms,
of cosmological scalar fields into black holes.

We shall make use of the fact that there exists a \emph{continuity
  equation} of the scalar field as long as the background spacetime is
static\cite{Choptuik:2003}. This fact will allow us to quantify the
absorption rate of a scalar wave packet by a black hole in a more
precise manner in terms of absorption flux and decay rates. For
simplicity, we will only focus our attention in the case of a massless
scalar field.

A brief summary of the paper is as follows. In Sec.~\ref{Eddington} we
set the mathematical background for the equations of motion, boundary
conditions, and initial conditions for the scalar field's wave
packet. Here we also show the existence of a continuity equation
arising directly from the equation of motion of the scalar field. In
Sec.~\ref{sec:numerical-results}, we present the main numerical
results, and the description of the fall of the scalar field in terms
of a normalized absorption rate. The latter arises naturally from the
use of the continuity equation found in Sec.~\ref{Eddington}. Finally,
Sec.~\ref{conclusions} is devoted to conclusions and final
comments. 

\section{\label{Eddington}Mathematical background}
We first consider a fixed Schwarzschild background with an
Eddintong-Finkelstein (EF) gauge, which is defined such that $t+r$ is
an ingoing null coordinate. Using the $3+1$ decomposition of the
metric\cite{Alcubierre08a,Thornburg:1998cx},  the $3$-metric
$\gamma_{ij}$ is
\begin{equation}
  \label{eq:metric}
  \gamma_{ij} = \mathrm{diag}\left[ a^2(r), r^2 , r^2 \sin^2 \theta
  \right] \, ,
\end{equation}
where $a^2(r) = 1 + 2GM/r$, $G$ is Newton's constant, and $M$ denotes
the mass of the black hole. The lapse $\alpha$ and shift
$\beta^i=[\beta,0,0]$ functions are, respectively,
\begin{equation}
  \label{gauge-quantities}
  \alpha(r) = a^{-1}(r) \, , \quad \beta(r) = \frac{2GM}{a^2(r)r} \, .
\end{equation} 
It is illustrative to calculate the coordinate velocities of null
geodesics, that are given by
\begin{equation}
  \label{eq:null}
  \frac{dr}{dt} = c_\pm \equiv - \beta \pm \alpha/a \, .
\end{equation}
Notice that the use of the EF gauge, from
Eqs.~(\ref{gauge-quantities}), is manifest through the condition $c_-
=-1$ for all points in the background spacetime (we use units in which
$c=1$).

The Klein-Gordon (KG) equation for a massless self-interacting scalar
field $\phi$ is 
\begin{equation}
  \frac{1}{\sqrt{-g}} \partial_\mu \left( \sqrt{-g} g^{\mu
      \nu} \partial_\nu \phi \right) = 0 \, . \label{klein-gordon}
\end{equation}
In order to solve it, it proves convenient to define two first order
variables\cite{Thornburg:1998cx,Alcubierre:2003sx},
\begin{equation}
 \Pi(t,r) = \partial_r \phi \, , \quad \eta(t,r) = \frac{1}{\alpha}
 \left( \partial_t \phi - \beta \Pi \right) \, , \label{pi-eta}
\end{equation}
with the help of which Eq.~(\ref{klein-gordon}) is represented by the
following three first order equations
\begin{subequations}
  \begin{eqnarray}
 \label{evolution-eq1}
    \frac{1}{\alpha} \left( \partial_t \phi - \beta \partial_r \phi
    \right) &=& \eta \, , \\ 
 \label{evolution-eq2}
     \frac{1}{\alpha} \left( \partial_t \Pi - \beta \partial_r \Pi
    \right) &=& \partial_r \eta + \frac{\beta}{2r} \eta + \partial_r
    \beta \, \Pi \, , \\ 
 \label{evolution-eq3}
    \frac{1}{\alpha} \left( \partial_t \eta - \beta \partial_r \eta
    \right) &=& \frac{\partial_r \Pi}{a^2} + \frac{a^2+1}{r}
    \frac{\Pi}{a^4} + K \eta \, ,
  \end{eqnarray}
\end{subequations}
where $K$ is the trace of the extrinsic
curvature. Eq.~(\ref{evolution-eq1}) arises from the very definition
of $\eta$ and $\Pi$, whereas the equation for $\Pi$,
Eq.~(\ref{evolution-eq2}), arises from the combination of
Eqs.~(\ref{pi-eta}); that of $\eta$ in Eq.~(\ref{evolution-eq3})
arises from the original KG equation~(\ref{klein-gordon}).

We shall take the quantity $(GM)$ as the unit for distance and time, so
that the radial and time coordinates are made dimensionless through
the change $r \to (GM) \hat{r}$ and $t \to (GM) \hat{t}$, where a hat
denotes dimensionless variables. Notice that the unit for distance and
time is half the usual Schwarzschild radius, $r_S \equiv 2GM$. The
scalar field is made dimensionless by the change $\phi \to \hat{\phi}
/\sqrt{G}$. Accordingly, the rest of the scalar field variables should
be changed by the expressions $\Pi \to (m^3_{Pl}/M) \hat{\Pi}$, and
$\eta \to (m^3_{Pl}/M) \hat{\eta}$, where the Planck mass is defined
as $m^2_{Pl} = G^{-1}$. 

We will impose outgoing-radiation boundary condition upon our
field variables at the outermost points of the numerical grid. As for
the innermost points, as long as they are inside the event horizon,
there is no need to put a boundary condition, because the light cones
there point inwards in the EF gauge, that is, $c_+ < 0$ and $c_- =-1$,
see Eqs.~(\ref{gauge-quantities}), and~(\ref{eq:null}).

The initial data for the scalar field in our numerical experiments
will be a Gaussian profile modulated by a spherical wave of the form
\begin{equation}
  \label{eq:initial}
  \hat{\phi}(r) = A \frac{\cos{( \hat{k}_0 \hat{r})}}{\hat{r}}
  e^{-\left( \hat{r} - \hat{r}_0 \right)^2/ \hat{\sigma}^2} \, ,  
\end{equation}
which is centered at $\hat{r} = \hat{r}_0$, and has amplitude $A$
and width $\hat{\sigma}$. The Gaussian distribution of wavenumbers in
Fourier space has a mean value $\langle \hat{k} \rangle = \hat{k}_0$,
and variance $\langle (\hat{k} - \hat{k}_0 )^2 \rangle =
1/\hat{\sigma}^2$. It should be noticed here that the wave number was
made dimensionless by the change $k \to \hat{k}/(GM)$.

On the other hand, by means of a lengthly but otherwise
straightforward calculation, it can be shown that the KG
equation~(\ref{klein-gordon}) can be written in the form of a
continuity equation\cite{Choptuik:2003}
\begin{equation}
 \label{eq:continuity}
 \partial_t \hat{\rho} - \frac{1}{\hat{r}^2} \partial_{\hat{r}}
 \left( \hat{r}^2 \hat{J}^r \right) = 0 \, ,
\end{equation}
where the charge density $\hat{\rho}$ and the scalar field current
density $\hat{J}^r$ are, respectively,
\begin{subequations}
\begin{eqnarray}
  \hat{\rho} &=& \frac{1}{2} \left( \hat{\eta}^2 +
  \frac{\hat{\Pi}^2}{a^2} \right) + \frac{\beta}{\alpha}
  \hat{\Pi}\hat{\eta} \, , \label{density-eq} \\
  \hat{J}^r &=& \beta \left( \hat{\eta}^2 + \alpha^2 \hat{\Pi}^2
  \right) + \alpha \left( \alpha^2 + a^2 \beta^2 \right) \hat{\Pi} \,
  \hat{\eta} \, . \label{flux-energy}
\end{eqnarray}
\end{subequations} 
In what follows, we will skip the hats of the variables, in the
understanding that they have been made dimensionless. We may use a hat
again on the variables whenever confusion may arise.

\section{\label{sec:numerical-results}Numerical results}
\begin{figure}[!tbp]
\includegraphics[width=0.49\textwidth]{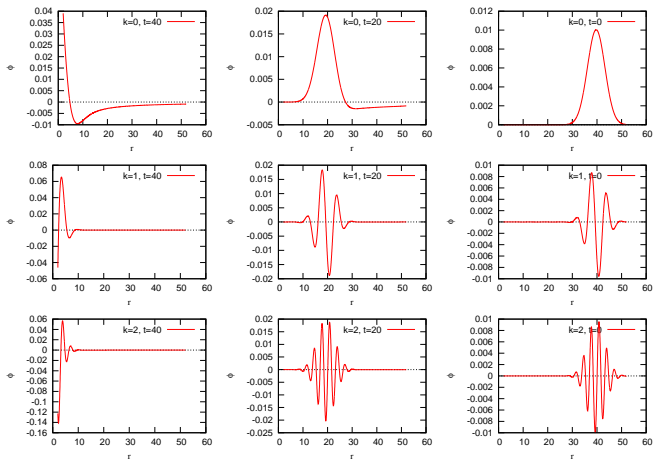}
\caption{\label{fig:figure1}Motion of a Gaussian packet for different
  values of its mean wavenumber $k_0$, see Eq.~(\ref{eq:initial}). The
  wave packets retain their shape as they approach the black hole's horizon,
  located at $r=2$. The wave packets were given ingoing initial
  conditions, $\eta = \Pi$. The time in each row proceeds from right
  to left.}
\end{figure}

The massless scalar field corresponds exactly to the solution of the
(homegeneous) wave equation in a curved spacetime. One important
feature of the massless case, which is helpful for the study of wave
packets, is that the field keeps its shape as it falls into the black
hole. Heuristically, this can be seen from the fact that the phase
$v_p$ and group $v_g$ velocities of the wave packet are both equal to
that of light, $v_p = v_g = c_\pm$. For illustration purposes, this
nice feature can be seen directly in the motion of the wave packets
shown in Fig.~\ref{fig:figure1}.

\begin{figure}[!tbp]
\includegraphics[width=0.5\textwidth]{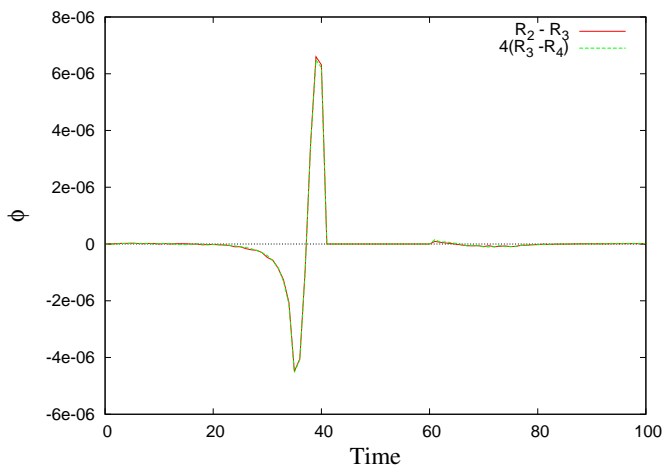}
\caption{\label{fig:figure13}Convergence plot for the numerical code
  used to solve the KG equation in a fixed Schwarzschild
  background. The example corresponds to the maximum of the wave
  function $\phi_{max}$ at each time in the simplest case $k_0
  =0$ (see also Fig.~\ref{fig:figure1}). Shown are three different
  runs with resolutions $R_1$: $\Delta r= 0.025$, $R_2$: $\Delta
  r=0.0125$, and $R_3$: $\Delta r=0.00625$. The plots, after the
  indicated scaling, agree in their profile, then the numerical output
  is second order convergent.
}
\end{figure}

We are solving differential equations with a finite differencing
method, which involves the truncation of a Taylor series expansion. It
is then necessary to show the proper convergence of numerical output
as the spatial grid is refined. In Fig.~\ref{fig:figure13}, we show
numerical runs with three different resolutions, where $R_1$ has
coarse resolution, $R_2$ has medium resolution, and $R_3$ is the
finest. As expected, the runs show that the numerical code is second
order convergent.

To study the rate at which the a scalar field wave packet is absorbed
by the black hole, we rely on the continuity equation. Notice that
Eq.~(\ref{eq:continuity}) looks pretty much the same as a
typical conservation equation in flat spacetime. Taking a bounded
proper volume $V$, we find
\begin{equation}
  \partial_t \int_V \sqrt{-g} \, \rho \, dr \, d\Omega = \partial_t
  \int_V 4\pi r^2 \, \rho \, dr = 4\pi \left. (r^2 J^r)
  \right|^{r_2}_{r_1} \, .
\end{equation}
Under the assumption that the scalar field current decays rapidly
enough as $r_2 \to \infty$, we find the useful result
\begin{equation}
  \frac{1}{M_\phi} \frac{dM_\phi}{dt} = - \frac{4\pi}{M_\phi}
  \left. (r^2 J^r) \right|_{r_S} = - \Gamma (t) \, ,
 \label{eq:Gamma}
\end{equation}
where $M_\phi = 4\pi \int_V r^2 \, \rho \, dr$ is the total scalar
field mass contained in the proper volume $V$. In fact, $M_\phi$ is
the conserved charge of the field as suggested by the continuity
equation~(\ref{eq:continuity}). We can monitor the absorption rate of
the (total) mass of the wave packet by calculating the scalar field
current going through the inner surface; in our case, the inner radius
is the black hole's horizon, $r_1 = r_S$.

Following standard notation in Physics, we have denoted the decay rate
of the wave packet's mass as $\Gamma (t)$, whose units are given in
terms of $(GM)^{-1}$. In our case, this decay rate is just the
normalized flux at the horizon of the black hole, being the
normalization factor the total mass of the wave packet that still
remains outside the black hole's horizon. 

\begin{figure}[!tbp]
\includegraphics[width=0.49\textwidth]{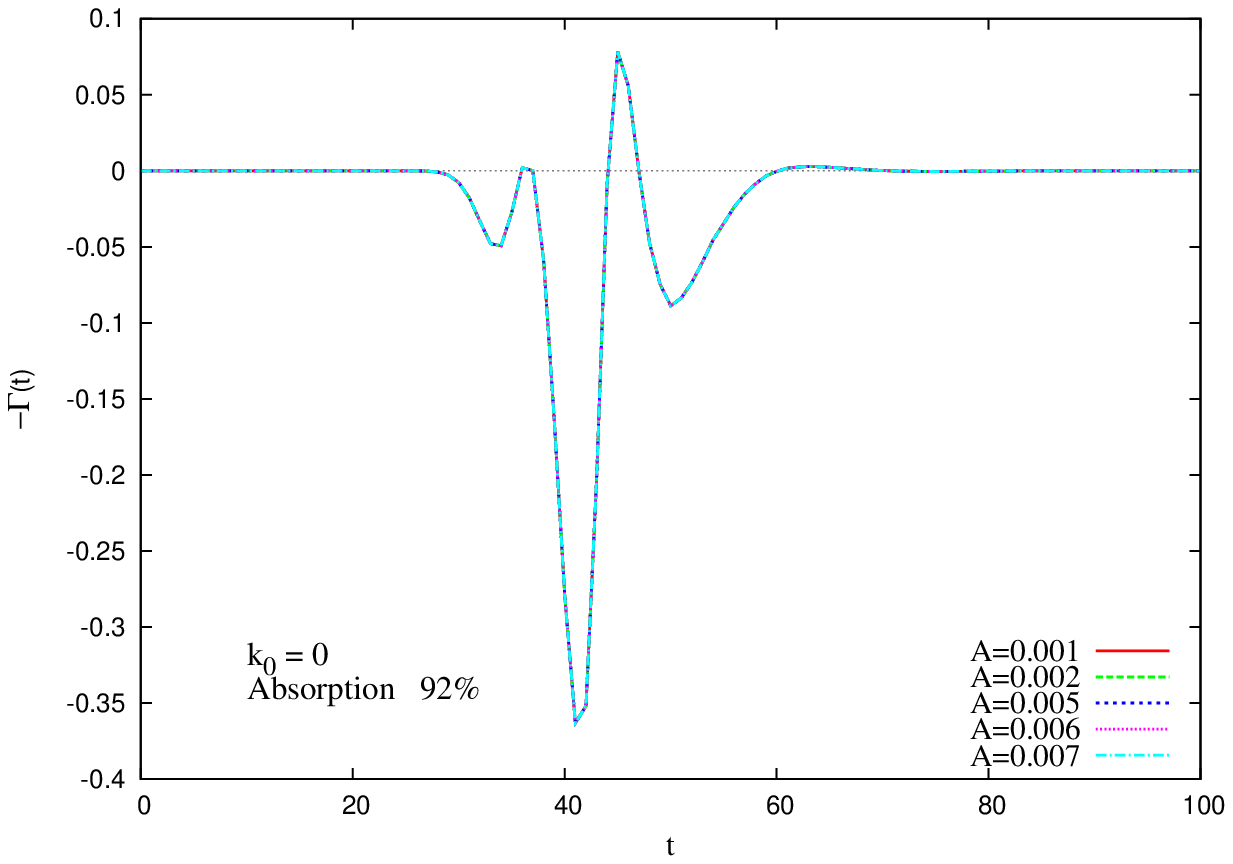}
\includegraphics[width=0.49\textwidth]{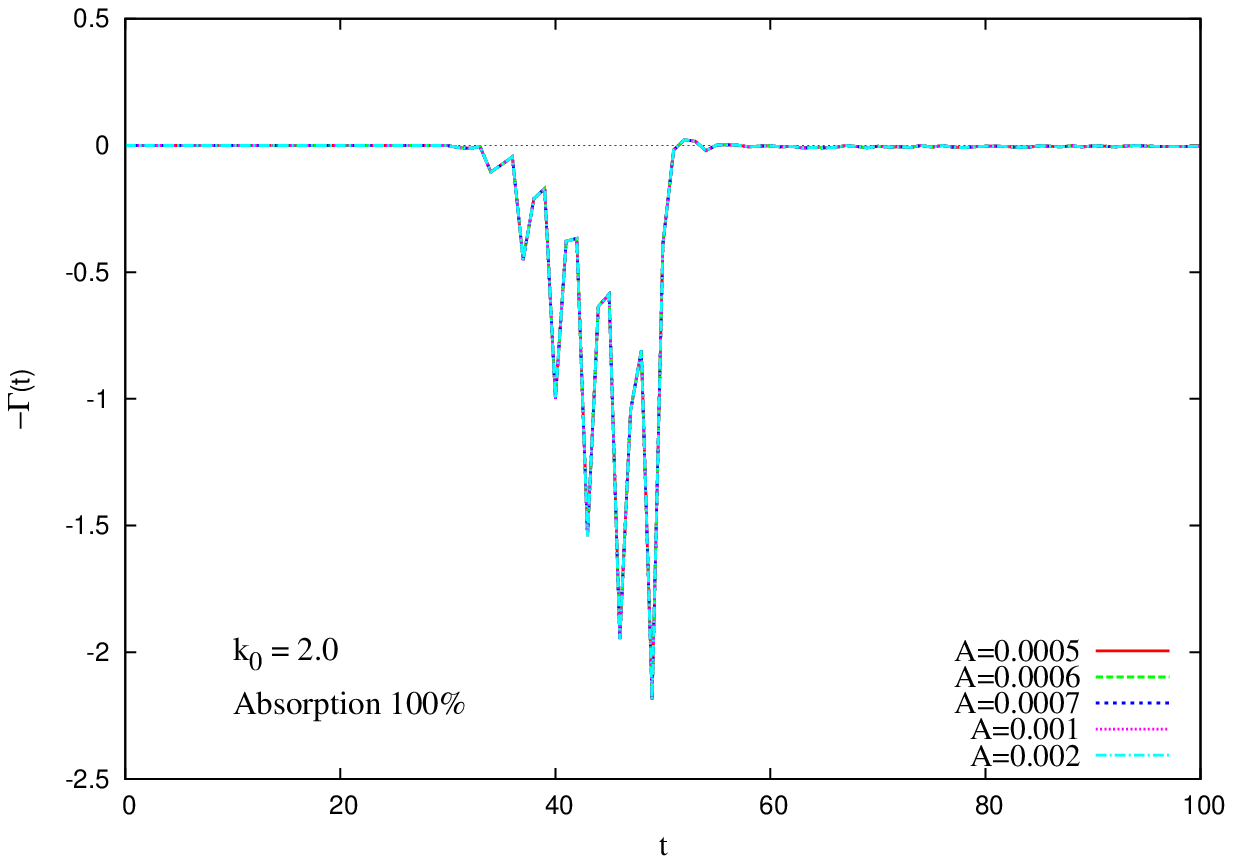}
\caption{The (normalized) decay rate $\Gamma(t)$, see
  Eq.~(\ref{eq:Gamma}), of a massless scalar field $\phi$ as it
  flows through the black hole's horizon. The units are given in terms
  of $(GM)^{-1}$, which for a typical mass of $10^6 M_\odot$
  gives $\Gamma \sim 0.2 \mathrm{s}^{-1}$. The runs were performed for
  different values of the amplitude $A$ and of the mean wavenumber $k_0$ of
  the wave packet~(\ref{eq:initial}), whereas the width was fixed to
  $\sigma = 5$. The total interaction time between the black hole and the
  wave packet is around $t_T \simeq 4\sigma$ in the two cases. The decay
  rate increases for larger values of $k_0$, but it is insensitive to
  the values of the amplitude $A$, i.e., to values of the scalar
  field's strength.
}
\label{fig:figure2}
\end{figure}

Typical curves for the decay rate are shown in
Fig.~\ref{fig:figure2}. An interesting and unexpected result is that
the (normalized) decay rate $\Gamma(t)$ does not depend on the
Gaussian's amplitude $A$, that is, it does not depend on the field's
strength. This means that larger packets are absorbed at the same rate as
are smaller packets. We can notice though that the mean wavenumber has an
effect on the absorption, as the latter increases for larger values of
$k_0$.

If we integrate Eq.~(\ref{eq:Gamma}), we can find the total mass
outside the horizon as a function of time,
\begin{equation}
 \frac{M_\phi (t)}{M_{\phi,i}} = \exp \left( -\int \Gamma(t) dt
 \right) = \exp \left( -\Delta(t)  \right) \, ,
\label{eq:absorbed}
\end{equation}
where $M_{\phi,i}$ is the initial total mass, and $\Delta(t)$ is the
(exponential) absorption ratio. If the integral is calculated for the
total time the wave packet is interacting with the black hole, it should
give us the total absorption ratio of the wave packet. For the
cases shown in Fig.~\ref{fig:figure2}, we have found that absorption
is about $92\%$ for $k_0 =0$, and $100\%$ for $k_0 = 2$. As a matter
of fact, our numerical experiments showed that total absorption is
always achieved if $k_0 \geq r^{-1}_S$ (see
also\cite{Hernandez:2007jt}).

\begin{figure}[!tbp]
\includegraphics[width=0.5\textwidth]{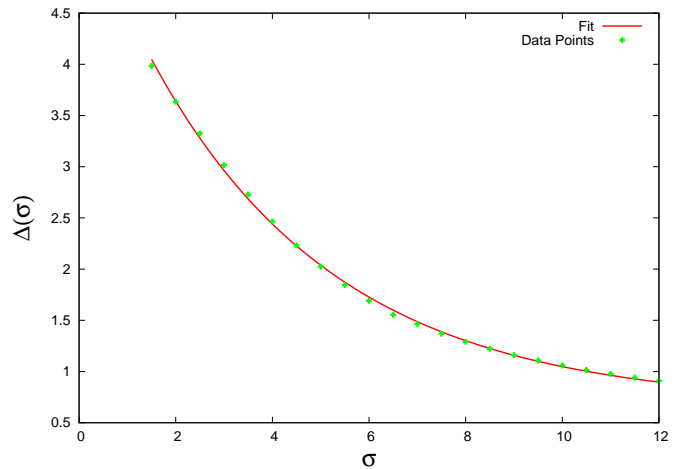}
\caption{Total absorption ratio $\Delta$, see
  Eq.~(\ref{eq:absorbed}), for different widths of the wave packets; for
  all runs $k_0 = 0$. The amplitude $A$ was adjust in each case as to
  have always the same total mass $M_\phi$ at the initial time. This
  is not strictly necessary, as the decay rate is independent of the
  packet's amplitude. We also show the fitting
  function~(\ref{eq:fit}).
}
\label{fig:figure3}
\end{figure}

As a final step, we study the dependence of $\Delta$ on the
width of the wave packet; for definiteness, we focus our attention in
the case $k_0=0$, which is also the most dispersive one. Numerical
results are shown in Fig.~\ref{fig:figure3}, and we notice that the
absorption decreases as the wave packet becomes wider. It can be
verified that the points can be fitted by a function of the form
\begin{equation}
  \label{eq:fit}
  \Delta(\sigma) = e_0 e^{-e_1 \sigma} + e_2 \, ,
\end{equation}
where, in the present case, a fitting procedure shows that $e_0=
4.98$, $e_1 = 0.259$, and $e_2 = 0.67$. In particular, if wave packets
as wide as necessary were allowed, then Eq.~(\ref{eq:fit}) suggests
that the total absorption would be
\begin{equation}
  \label{eq:continous}
  \lim_{\sigma \to \infty} \Delta(\sigma) = 0.67 \, .
\end{equation}

\section{\label{conclusions} Conclusions}
The motion of a wave packet in a black hole spacetime raised some
interest in the cosmological community because of the possibility that
SMBH could have grown because of the accretion a Quintessence-type
scalar field. This is not the only possible case, but we can ask the
same question about any other cosmological scalar field living around
a black hole.

We have explored the simplest possibility, that of a massless scalar
field, for the motion of a wave packet in a fixed black hole spacetime
using an EF gauge. To have a better visualization of the absorption of
the scalar field by the black hole, we took advantage of the fact that
one can write out a continuity equation from the KG equation. 

The corresponding conserved charge is the total mass of the wave
packet, but more important is the definition of a current density,
with the help of which we were able to define a (normalized) decay
rate for the wave packet, whose magnitude is given by the black
hole's mass, $\Gamma(t) \sim (GM)^{-1}$. This means that less massive
black holes accrete scalar field matter at a larger rate; for example,
a black hole as massive as the Sun would accrete at an incredible rate
of $\Gamma \sim 10^5\mathrm{s}^{-1}$! In terms of the decay rate,
we too found that the absorption depends on the mean wavenumber of the
wave packet; actually, full absorption is reached for mean wavelengths
smaller than the Schwarzschild radius, $k_0 < r^{-1}_S$.

However, a new result showed up: the decay rate does not depend on the
scalar field's strength. Moreover, we could use this result to show
the dependence of the absorption on the packet's width. By a limiting
procedure on a fitting function, we determined the maximum total
absorption of a wave packet with a width much larger than the black
hole's horizon: around $e^{-0.67} \simeq 0.51$. This is the result that
may have relevance for cosmology, as we expect cosmological scalar
fields to have very large intrinsic length scales ($k_0 \to 0$ and
$\sigma \to \infty$) as compared to the Schwarzschild radius of
supermassive black holes.

In the massless case studied here, there were two length scales
involved: the mean wavelength $\lambda_0 = k^{-1}_0$, and the width of
the wave packet $\sigma$. We were able to show the general dependence
of the (normalized) absorption rate on these length scales. In
general, we can say that black holes are quite efficient in absorbing
scalar fields, even in the case of very wide packets.

The method outlined here can be extended to the massive case. However,
in the case the scalar field has a mass $m$, additional length scales
appears in the problem in the form of the Compton length of the field,
$\lambda_ C \equiv m^{-1}$ and the Schwarzschild radius $r_S$ (in the
massless case, the Schwarzschild radius does not appear explicitly in
the equations of motion) that may introduce non-trivial features in
the motion of the wave packet and its absorption rate. This is ongoing
research that we expect to report elsewhere.

\begin{acknowledgments}
We are grateful to Francisco S. Guzm\'an for useful comments and help on
the numerical implementation of our work. LMF acknowledges support
from CONACyT, M\'exico. LAU-L thanks the Berkeley Center for
Cosmological Physics (BCCP) for its kind hospitality, and the joint
support of the Academia Mexiana de Ciencias and the United
States-Mexico Foundation for Science for a summer research stay at
BCCP. This work was partially supported by PROMEP, DAIP, and by
CONACyT M\'exico under grants 56946, and I0101/131/07 C-234/07 of the
Instituto Avanzado de Cosmologia (IAC) collaboration
(http://www.iac.edu.mx/).
\end{acknowledgments}

\bibliography{black-ref}

\begin{thebibliography}{41}%
\makeatletter
\providecommand \@ifxundefined [1]{%
 \@ifx{#1\undefined}
}%
\providecommand \@ifnum [1]{%
 \ifnum #1\expandafter \@firstoftwo
 \else \expandafter \@secondoftwo
 \fi
}%
\providecommand \@ifx [1]{%
 \ifx #1\expandafter \@firstoftwo
 \else \expandafter \@secondoftwo
 \fi
}%
\providecommand \natexlab [1]{#1}%
\providecommand \enquote  [1]{``#1''}%
\providecommand \bibnamefont  [1]{#1}%
\providecommand \bibfnamefont [1]{#1}%
\providecommand \citenamefont [1]{#1}%
\providecommand \href@noop [0]{\@secondoftwo}%
\providecommand \href [0]{\begingroup \@sanitize@url \@href}%
\providecommand \@href[1]{\@@startlink{#1}\@@href}%
\providecommand \@@href[1]{\endgroup#1\@@endlink}%
\providecommand \@sanitize@url [0]{\catcode `\\12\catcode `\$12\catcode
  `\&12\catcode `\#12\catcode `\^12\catcode `\_12\catcode `\%12\relax}%
\providecommand \@@startlink[1]{}%
\providecommand \@@endlink[0]{}%
\providecommand \url  [0]{\begingroup\@sanitize@url \@url }%
\providecommand \@url [1]{\endgroup\@href {#1}{\urlprefix }}%
\providecommand \urlprefix  [0]{URL }%
\providecommand \Eprint [0]{\href }%
\@ifxundefined \urlstyle {%
  \providecommand \doi  [0]{\begingroup \@sanitize@url \@doi}%
  \providecommand \@doi [1]{\endgroup \@@startlink {\doibase
  #1}doi:\discretionary {}{}{}#1\@@endlink }%
}{%
  \providecommand \doi  [0]{doi:\discretionary{}{}{}\begingroup
  \urlstyle{rm}\Url }%
}%
\providecommand \doibase [0]{http://dx.doi.org/}%
\providecommand \Doi [0]{\begingroup \@sanitize@url \@Doi }%
\providecommand \@Doi  [1]{\endgroup\@@startlink{\doibase#1}\@@Doi}%
\providecommand \@@Doi [1]{#1\@@endlink}%
\providecommand \selectlanguage [0]{\@gobble}%
\providecommand \bibinfo  [0]{\@secondoftwo}%
\providecommand \bibfield  [0]{\@secondoftwo}%
\providecommand \translation [1]{[#1]}%
\providecommand \BibitemOpen [0]{}%
\providecommand \bibitemStop [0]{}%
\providecommand \bibitemNoStop [0]{.\EOS\space}%
\providecommand \EOS [0]{\spacefactor3000\relax}%
\providecommand \BibitemShut  [1]{\csname bibitem#1\endcsname}%
\bibitem [{\citenamefont {Caramete}\ and\ \citenamefont
  {Biermann}(2011)}]{Caramete:2011eh}%
  \BibitemOpen
  \bibfield  {author} {\bibinfo {author} {\bibfnamefont {L.~I.}\ \bibnamefont
  {Caramete}}\ and\ \bibinfo {author} {\bibfnamefont {P.~L.}\ \bibnamefont
  {Biermann}},\ }\href@noop {} { (\bibinfo {year} {2011})},\ \Eprint
  {http://arxiv.org/abs/1107.2244} {arXiv:1107.2244 [astro-ph.GA]} \BibitemShut
  {NoStop}%
\bibitem [{\citenamefont {Volonteri}\ and\ \citenamefont
  {Stark}(2011)}]{Volonteri:2011rm}%
  \BibitemOpen
  \bibfield  {author} {\bibinfo {author} {\bibfnamefont {M.}~\bibnamefont
  {Volonteri}}\ and\ \bibinfo {author} {\bibfnamefont {D.~P.}\ \bibnamefont
  {Stark}},\ }\href@noop {} { (\bibinfo {year} {2011})},\ \Eprint
  {http://arxiv.org/abs/1107.1946} {arXiv:1107.1946 [astro-ph.CO]} \BibitemShut
  {NoStop}%
\bibitem [{\citenamefont {Ferrarese}(2002)}]{Ferrarese:2002ct}%
  \BibitemOpen
  \bibfield  {author} {\bibinfo {author} {\bibfnamefont {L.}~\bibnamefont
  {Ferrarese}},\ }\href@noop {} {\bibfield  {journal} {\bibinfo  {journal}
  {Astrophys. J.},\ }\textbf {\bibinfo {volume} {578}},\ \bibinfo {pages} {90}
  (\bibinfo {year} {2002})},\ \Eprint {http://arxiv.org/abs/astro-ph/0203469}
  {astro-ph/0203469} \BibitemShut {NoStop}%
\bibitem [{\citenamefont {Nowak}\ \emph {et~al.}(2007)\citenamefont {Nowak}
  \emph {et~al.}}]{Nowak:2007ba}%
  \BibitemOpen
  \bibfield  {author} {\bibinfo {author} {\bibfnamefont {N.}~\bibnamefont
  {Nowak}} \emph {et~al.},\ }\Doi {10.1111/j.1365-2966.2007.11949.x} {\bibfield
   {journal} {\bibinfo  {journal} {Mon. Not. Roy. Astron. Soc.},\ }\textbf
  {\bibinfo {volume} {379}},\ \bibinfo {pages} {909} (\bibinfo {year}
  {2007})},\ \Eprint {http://arxiv.org/abs/0705.1758} {arXiv:0705.1758
  [astro-ph]} \BibitemShut {NoStop}%
\bibitem [{\citenamefont {Greenwood}(2005)}]{Greenwood:2005cs}%
  \BibitemOpen
  \bibfield  {author} {\bibinfo {author} {\bibfnamefont {C.~J.}\ \bibnamefont
  {Greenwood}},\ }\href@noop {} { (\bibinfo {year} {2005})},\ \Eprint
  {http://arxiv.org/abs/astro-ph/0512350} {arXiv:astro-ph/0512350} \BibitemShut
  {NoStop}%
\bibitem [{\citenamefont {Ghez}\ \emph {et~al.}(2000)\citenamefont {Ghez},
  \citenamefont {Morris}, \citenamefont {Becklin}, \citenamefont {Kremenek},\
  and\ \citenamefont {Tanner}}]{Ghez:2000ay}%
  \BibitemOpen
  \bibfield  {author} {\bibinfo {author} {\bibfnamefont {A.}~\bibnamefont
  {Ghez}}, \bibinfo {author} {\bibfnamefont {M.}~\bibnamefont {Morris}},
  \bibinfo {author} {\bibfnamefont {E.~E.}\ \bibnamefont {Becklin}}, \bibinfo
  {author} {\bibfnamefont {T.}~\bibnamefont {Kremenek}}, \ and\ \bibinfo
  {author} {\bibfnamefont {A.}~\bibnamefont {Tanner}},\ }\href@noop {}
  {\bibfield  {journal} {\bibinfo  {journal} {Nature},\ }\textbf {\bibinfo
  {volume} {407}},\ \bibinfo {pages} {349} (\bibinfo {year} {2000})},\ \Eprint
  {http://arxiv.org/abs/astro-ph/0009339} {astro-ph/0009339} \BibitemShut
  {NoStop}%
\bibitem [{\citenamefont {Melia}(2007)}]{Melia:2007vt}%
  \BibitemOpen
  \bibfield  {author} {\bibinfo {author} {\bibfnamefont {F.}~\bibnamefont
  {Melia}},\ }\href@noop {} { (\bibinfo {year} {2007})},\ \Eprint
  {http://arxiv.org/abs/0705.1537} {arXiv:0705.1537 [astro-ph]} \BibitemShut
  {NoStop}%
\bibitem [{\citenamefont {Carr}\ \emph
  {et~al.}(2010){\natexlab{a}}\citenamefont {Carr}, \citenamefont {Kohri},
  \citenamefont {Sendouda},\ and\ \citenamefont {Yokoyama}}]{Carr:2009jm}%
  \BibitemOpen
  \bibfield  {author} {\bibinfo {author} {\bibfnamefont {B.}~\bibnamefont
  {Carr}}, \bibinfo {author} {\bibfnamefont {K.}~\bibnamefont {Kohri}},
  \bibinfo {author} {\bibfnamefont {Y.}~\bibnamefont {Sendouda}}, \ and\
  \bibinfo {author} {\bibfnamefont {J.}~\bibnamefont {Yokoyama}},\ }\Doi
  {10.1103/PhysRevD.81.104019} {\bibfield  {journal} {\bibinfo  {journal}
  {Phys.Rev.},\ }\textbf {\bibinfo {volume} {D81}},\ \bibinfo {pages} {104019}
  (\bibinfo {year} {2010}{\natexlab{a}})},\ \Eprint
  {http://arxiv.org/abs/0912.5297} {arXiv:0912.5297 [astro-ph.CO]} \BibitemShut
  {NoStop}%
\bibitem [{\citenamefont {Bean}\ and\ \citenamefont
  {Magueijo}(2002)}]{Bean:2002kx}%
  \BibitemOpen
  \bibfield  {author} {\bibinfo {author} {\bibfnamefont {R.}~\bibnamefont
  {Bean}}\ and\ \bibinfo {author} {\bibfnamefont {J.}~\bibnamefont
  {Magueijo}},\ }\href@noop {} {\bibfield  {journal} {\bibinfo  {journal}
  {Phys. Rev.},\ }\textbf {\bibinfo {volume} {D66}},\ \bibinfo {pages} {063505}
  (\bibinfo {year} {2002})},\ \Eprint {http://arxiv.org/abs/astro-ph/0204486}
  {astro-ph/0204486} \BibitemShut {NoStop}%
\bibitem [{\citenamefont {Babichev}\ \emph {et~al.}(2004)\citenamefont
  {Babichev}, \citenamefont {Dokuchaev},\ and\ \citenamefont
  {Eroshenko}}]{Babichev:2004yx}%
  \BibitemOpen
  \bibfield  {author} {\bibinfo {author} {\bibfnamefont {E.}~\bibnamefont
  {Babichev}}, \bibinfo {author} {\bibfnamefont {V.}~\bibnamefont {Dokuchaev}},
  \ and\ \bibinfo {author} {\bibfnamefont {Y.}~\bibnamefont {Eroshenko}},\
  }\Doi {10.1103/PhysRevLett.93.021102} {\bibfield  {journal} {\bibinfo
  {journal} {Phys.Rev.Lett.},\ }\textbf {\bibinfo {volume} {93}},\ \bibinfo
  {pages} {021102} (\bibinfo {year} {2004})},\ \Eprint
  {http://arxiv.org/abs/gr-qc/0402089} {arXiv:gr-qc/0402089 [gr-qc]}
  \BibitemShut {NoStop}%
\bibitem [{\citenamefont {Babichev}\ \emph {et~al.}(2005)\citenamefont
  {Babichev}, \citenamefont {Dokuchaev},\ and\ \citenamefont
  {Eroshenko}}]{Babichev:2005py}%
  \BibitemOpen
  \bibfield  {author} {\bibinfo {author} {\bibfnamefont {E.}~\bibnamefont
  {Babichev}}, \bibinfo {author} {\bibfnamefont {V.}~\bibnamefont {Dokuchaev}},
  \ and\ \bibinfo {author} {\bibfnamefont {Y.}~\bibnamefont {Eroshenko}},\
  }\Doi {10.1134/1.1901765, 10.1134/1.1901765} {\bibfield  {journal} {\bibinfo
  {journal} {J.Exp.Theor.Phys.},\ }\textbf {\bibinfo {volume} {100}},\ \bibinfo
  {pages} {528} (\bibinfo {year} {2005})},\ \Eprint
  {http://arxiv.org/abs/astro-ph/0505618} {arXiv:astro-ph/0505618 [astro-ph]}
  \BibitemShut {NoStop}%
\bibitem [{\citenamefont {Mersini-Houghton}\ and\ \citenamefont
  {Kelleher}(2008)}]{MersiniHoughton:2008aw}%
  \BibitemOpen
  \bibfield  {author} {\bibinfo {author} {\bibfnamefont {L.}~\bibnamefont
  {Mersini-Houghton}}\ and\ \bibinfo {author} {\bibfnamefont {A.}~\bibnamefont
  {Kelleher}},\ }\href@noop {} { (\bibinfo {year} {2008})},\ \Eprint
  {http://arxiv.org/abs/0808.3419} {arXiv:0808.3419 [gr-qc]} \BibitemShut
  {NoStop}%
\bibitem [{\citenamefont {Rodrigues}\ and\ \citenamefont
  {Saa}(2009)}]{Rodrigues:2009eg}%
  \BibitemOpen
  \bibfield  {author} {\bibinfo {author} {\bibfnamefont {M.~G.}\ \bibnamefont
  {Rodrigues}}\ and\ \bibinfo {author} {\bibfnamefont {A.}~\bibnamefont
  {Saa}},\ }\Doi {10.1103/PhysRevD.80.104018} {\bibfield  {journal} {\bibinfo
  {journal} {Phys.Rev.},\ }\textbf {\bibinfo {volume} {D80}},\ \bibinfo {pages}
  {104018} (\bibinfo {year} {2009})},\ \Eprint {http://arxiv.org/abs/0909.3033}
  {arXiv:0909.3033 [gr-qc]} \BibitemShut {NoStop}%
\bibitem [{\citenamefont {Jacobson}(1999)}]{Jacobson:1999vr}%
  \BibitemOpen
  \bibfield  {author} {\bibinfo {author} {\bibfnamefont {T.}~\bibnamefont
  {Jacobson}},\ }\href@noop {} {\bibfield  {journal} {\bibinfo  {journal}
  {Phys. Rev. Lett.},\ }\textbf {\bibinfo {volume} {83}},\ \bibinfo {pages}
  {2699} (\bibinfo {year} {1999})},\ \Eprint
  {http://arxiv.org/abs/astro-ph/9905303} {astro-ph/9905303} \BibitemShut
  {NoStop}%
\bibitem [{\citenamefont {Frolov}\ and\ \citenamefont
  {Kofman}(2003)}]{Frolov:2002va}%
  \BibitemOpen
  \bibfield  {author} {\bibinfo {author} {\bibfnamefont {A.~V.}\ \bibnamefont
  {Frolov}}\ and\ \bibinfo {author} {\bibfnamefont {L.}~\bibnamefont
  {Kofman}},\ }\href@noop {} {\bibfield  {journal} {\bibinfo  {journal}
  {JCAP},\ }\textbf {\bibinfo {volume} {0305}},\ \bibinfo {pages} {009}
  (\bibinfo {year} {2003})},\ \Eprint {http://arxiv.org/abs/hep-th/0212327}
  {hep-th/0212327} \BibitemShut {NoStop}%
\bibitem [{\citenamefont {Harada}\ and\ \citenamefont
  {Carr}(2005)}]{Harada:2004pf}%
  \BibitemOpen
  \bibfield  {author} {\bibinfo {author} {\bibfnamefont {T.}~\bibnamefont
  {Harada}}\ and\ \bibinfo {author} {\bibfnamefont {B.~J.}\ \bibnamefont
  {Carr}},\ }\Doi {10.1103/PhysRevD.71.104010} {\bibfield  {journal} {\bibinfo
  {journal} {Phys.Rev.},\ }\textbf {\bibinfo {volume} {D71}},\ \bibinfo {pages}
  {104010} (\bibinfo {year} {2005})},\ \Eprint
  {http://arxiv.org/abs/astro-ph/0412135} {arXiv:astro-ph/0412135 [astro-ph]}
  \BibitemShut {NoStop}%
\bibitem [{\citenamefont {Custodio}\ and\ \citenamefont
  {Horvath}(2005)}]{Custodio:2005en}%
  \BibitemOpen
  \bibfield  {author} {\bibinfo {author} {\bibfnamefont {P.~S.}\ \bibnamefont
  {Custodio}}\ and\ \bibinfo {author} {\bibfnamefont {J.~E.}\ \bibnamefont
  {Horvath}},\ }\href@noop {} {\bibfield  {journal} {\bibinfo  {journal} {Int.
  J. Mod. Phys.},\ }\textbf {\bibinfo {volume} {D14}},\ \bibinfo {pages} {257}
  (\bibinfo {year} {2005})},\ \Eprint {http://arxiv.org/abs/gr-qc/0502118}
  {gr-qc/0502118} \BibitemShut {NoStop}%
\bibitem [{\citenamefont {Carr}\ \emph
  {et~al.}(2010){\natexlab{b}}\citenamefont {Carr}, \citenamefont {Harada},\
  and\ \citenamefont {Maeda}}]{Carr:2010wk}%
  \BibitemOpen
  \bibfield  {author} {\bibinfo {author} {\bibfnamefont {B.}~\bibnamefont
  {Carr}}, \bibinfo {author} {\bibfnamefont {T.}~\bibnamefont {Harada}}, \ and\
  \bibinfo {author} {\bibfnamefont {H.}~\bibnamefont {Maeda}},\ }\Doi
  {10.1088/0264-9381/27/18/183101} {\bibfield  {journal} {\bibinfo  {journal}
  {Class.Quant.Grav.},\ }\textbf {\bibinfo {volume} {27}},\ \bibinfo {pages}
  {183101} (\bibinfo {year} {2010}{\natexlab{b}})},\ \bibinfo {note} {*
  Temporary entry *},\ \Eprint {http://arxiv.org/abs/1003.3324}
  {arXiv:1003.3324 [gr-qc]} \BibitemShut {NoStop}%
\bibitem [{\citenamefont {Sahni}\ and\ \citenamefont
  {Wang}(2000)}]{Sahni:1999qe}%
  \BibitemOpen
  \bibfield  {author} {\bibinfo {author} {\bibfnamefont {V.}~\bibnamefont
  {Sahni}}\ and\ \bibinfo {author} {\bibfnamefont {L.-M.}\ \bibnamefont
  {Wang}},\ }\Doi {10.1103/PhysRevD.62.103517} {\bibfield  {journal} {\bibinfo
  {journal} {Phys.Rev.},\ }\textbf {\bibinfo {volume} {D62}},\ \bibinfo {pages}
  {103517} (\bibinfo {year} {2000})},\ \Eprint
  {http://arxiv.org/abs/astro-ph/9910097} {arXiv:astro-ph/9910097 [astro-ph]}
  \BibitemShut {NoStop}%
\bibitem [{\citenamefont {Matos}\ and\ \citenamefont
  {Urena-Lopez}(2000)}]{Matos:2000ng}%
  \BibitemOpen
  \bibfield  {author} {\bibinfo {author} {\bibfnamefont {T.}~\bibnamefont
  {Matos}}\ and\ \bibinfo {author} {\bibfnamefont {L.}~\bibnamefont
  {Urena-Lopez}},\ }\Doi {10.1088/0264-9381/17/13/101} {\bibfield  {journal}
  {\bibinfo  {journal} {Class.Quant.Grav.},\ }\textbf {\bibinfo {volume}
  {17}},\ \bibinfo {pages} {L75} (\bibinfo {year} {2000})},\ \Eprint
  {http://arxiv.org/abs/astro-ph/0004332} {arXiv:astro-ph/0004332 [astro-ph]}
  \BibitemShut {NoStop}%
\bibitem [{\citenamefont {Matos}\ and\ \citenamefont
  {Urena-Lopez}(2001)}]{Matos:2000ss}%
  \BibitemOpen
  \bibfield  {author} {\bibinfo {author} {\bibfnamefont {T.}~\bibnamefont
  {Matos}}\ and\ \bibinfo {author} {\bibfnamefont {L.~A.}\ \bibnamefont
  {Urena-Lopez}},\ }\Doi {10.1103/PhysRevD.63.063506} {\bibfield  {journal}
  {\bibinfo  {journal} {Phys.Rev.},\ }\textbf {\bibinfo {volume} {D63}},\
  \bibinfo {pages} {063506} (\bibinfo {year} {2001})},\ \Eprint
  {http://arxiv.org/abs/astro-ph/0006024} {arXiv:astro-ph/0006024 [astro-ph]}
  \BibitemShut {NoStop}%
\bibitem [{\citenamefont {Arbey}\ \emph {et~al.}(2001)\citenamefont {Arbey},
  \citenamefont {Lesgourgues},\ and\ \citenamefont {Salati}}]{Arbey:2001qi}%
  \BibitemOpen
  \bibfield  {author} {\bibinfo {author} {\bibfnamefont {A.}~\bibnamefont
  {Arbey}}, \bibinfo {author} {\bibfnamefont {J.}~\bibnamefont {Lesgourgues}},
  \ and\ \bibinfo {author} {\bibfnamefont {P.}~\bibnamefont {Salati}},\ }\Doi
  {10.1103/PhysRevD.64.123528} {\bibfield  {journal} {\bibinfo  {journal}
  {Phys.Rev.},\ }\textbf {\bibinfo {volume} {D64}},\ \bibinfo {pages} {123528}
  (\bibinfo {year} {2001})},\ \Eprint {http://arxiv.org/abs/astro-ph/0105564}
  {arXiv:astro-ph/0105564 [astro-ph]} \BibitemShut {NoStop}%
\bibitem [{\citenamefont {Matos}\ \emph {et~al.}(2008)\citenamefont {Matos},
  \citenamefont {Vazquez},\ and\ \citenamefont {Magana}}]{Matos:2008ag}%
  \BibitemOpen
  \bibfield  {author} {\bibinfo {author} {\bibfnamefont {T.}~\bibnamefont
  {Matos}}, \bibinfo {author} {\bibfnamefont {J.}~\bibnamefont {Vazquez}}, \
  and\ \bibinfo {author} {\bibfnamefont {J.}~\bibnamefont {Magana}},\
  }\href@noop {} { (\bibinfo {year} {2008})},\ \bibinfo {note} {* Brief entry
  *},\ \Eprint {http://arxiv.org/abs/0806.0683} {arXiv:0806.0683 [astro-ph]}
  \BibitemShut {NoStop}%
\bibitem [{\citenamefont {Arbey}(2006)}]{Arbey:2006it}%
  \BibitemOpen
  \bibfield  {author} {\bibinfo {author} {\bibfnamefont {A.}~\bibnamefont
  {Arbey}},\ }\Doi {10.1103/PhysRevD.74.043516} {\bibfield  {journal} {\bibinfo
   {journal} {Phys.Rev.},\ }\textbf {\bibinfo {volume} {D74}},\ \bibinfo
  {pages} {043516} (\bibinfo {year} {2006})},\ \Eprint
  {http://arxiv.org/abs/astro-ph/0601274} {arXiv:astro-ph/0601274 [astro-ph]}
  \BibitemShut {NoStop}%
\bibitem [{\citenamefont {Liddle}\ and\ \citenamefont
  {Urena-Lopez}(2006)}]{Liddle:2006qz}%
  \BibitemOpen
  \bibfield  {author} {\bibinfo {author} {\bibfnamefont {A.~R.}\ \bibnamefont
  {Liddle}}\ and\ \bibinfo {author} {\bibfnamefont {L.~A.}\ \bibnamefont
  {Urena-Lopez}},\ }\Doi {10.1103/PhysRevLett.97.161301} {\bibfield  {journal}
  {\bibinfo  {journal} {Phys.Rev.Lett.},\ }\textbf {\bibinfo {volume} {97}},\
  \bibinfo {pages} {161301} (\bibinfo {year} {2006})},\ \Eprint
  {http://arxiv.org/abs/astro-ph/0605205} {arXiv:astro-ph/0605205 [astro-ph]}
  \BibitemShut {NoStop}%
\bibitem [{\citenamefont {Sin}(1994)}]{Sin:1992bg}%
  \BibitemOpen
  \bibfield  {author} {\bibinfo {author} {\bibfnamefont {S.-J.}\ \bibnamefont
  {Sin}},\ }\Doi {10.1103/PhysRevD.50.3650} {\bibfield  {journal} {\bibinfo
  {journal} {Phys.Rev.},\ }\textbf {\bibinfo {volume} {D50}},\ \bibinfo {pages}
  {3650} (\bibinfo {year} {1994})},\ \Eprint
  {http://arxiv.org/abs/hep-ph/9205208} {arXiv:hep-ph/9205208 [hep-ph]}
  \BibitemShut {NoStop}%
\bibitem [{\citenamefont {Ji}\ and\ \citenamefont {Sin}(1994)}]{Ji:1994xh}%
  \BibitemOpen
  \bibfield  {author} {\bibinfo {author} {\bibfnamefont {S.}~\bibnamefont
  {Ji}}\ and\ \bibinfo {author} {\bibfnamefont {S.}~\bibnamefont {Sin}},\ }\Doi
  {10.1103/PhysRevD.50.3655} {\bibfield  {journal} {\bibinfo  {journal}
  {Phys.Rev.},\ }\textbf {\bibinfo {volume} {D50}},\ \bibinfo {pages} {3655}
  (\bibinfo {year} {1994})},\ \Eprint {http://arxiv.org/abs/hep-ph/9409267}
  {arXiv:hep-ph/9409267 [hep-ph]} \BibitemShut {NoStop}%
\bibitem [{\citenamefont {Arbey}\ \emph {et~al.}(2003)\citenamefont {Arbey},
  \citenamefont {Lesgourgues},\ and\ \citenamefont {Salati}}]{Arbey:2003sj}%
  \BibitemOpen
  \bibfield  {author} {\bibinfo {author} {\bibfnamefont {A.}~\bibnamefont
  {Arbey}}, \bibinfo {author} {\bibfnamefont {J.}~\bibnamefont {Lesgourgues}},
  \ and\ \bibinfo {author} {\bibfnamefont {P.}~\bibnamefont {Salati}},\ }\Doi
  {10.1103/PhysRevD.68.023511} {\bibfield  {journal} {\bibinfo  {journal}
  {Phys.Rev.},\ }\textbf {\bibinfo {volume} {D68}},\ \bibinfo {pages} {023511}
  (\bibinfo {year} {2003})},\ \Eprint {http://arxiv.org/abs/astro-ph/0301533}
  {arXiv:astro-ph/0301533 [astro-ph]} \BibitemShut {NoStop}%
\bibitem [{\citenamefont {Alcubierre}\ \emph {et~al.}(2003)\citenamefont
  {Alcubierre}, \citenamefont {Becerril}, \citenamefont {Guzman}, \citenamefont
  {Matos}, \citenamefont {Nunez} \emph {et~al.}}]{Alcubierre:2003sx}%
  \BibitemOpen
  \bibfield  {author} {\bibinfo {author} {\bibfnamefont {M.}~\bibnamefont
  {Alcubierre}}, \bibinfo {author} {\bibfnamefont {R.}~\bibnamefont
  {Becerril}}, \bibinfo {author} {\bibfnamefont {S.~F.}\ \bibnamefont
  {Guzman}}, \bibinfo {author} {\bibfnamefont {T.}~\bibnamefont {Matos}},
  \bibinfo {author} {\bibfnamefont {D.}~\bibnamefont {Nunez}},  \emph
  {et~al.},\ }\Doi {10.1088/0264-9381/20/13/332} {\bibfield  {journal}
  {\bibinfo  {journal} {Class.Quant.Grav.},\ }\textbf {\bibinfo {volume}
  {20}},\ \bibinfo {pages} {2883} (\bibinfo {year} {2003})},\ \Eprint
  {http://arxiv.org/abs/gr-qc/0301105} {arXiv:gr-qc/0301105 [gr-qc]}
  \BibitemShut {NoStop}%
\bibitem [{\citenamefont {Guzman}\ and\ \citenamefont
  {Urena-Lopez}(2003)}]{Guzman:2003kt}%
  \BibitemOpen
  \bibfield  {author} {\bibinfo {author} {\bibfnamefont {F.}~\bibnamefont
  {Guzman}}\ and\ \bibinfo {author} {\bibfnamefont {L.~A.}\ \bibnamefont
  {Urena-Lopez}},\ }\Doi {10.1103/PhysRevD.68.024023} {\bibfield  {journal}
  {\bibinfo  {journal} {Phys.Rev.},\ }\textbf {\bibinfo {volume} {D68}},\
  \bibinfo {pages} {024023} (\bibinfo {year} {2003})},\ \Eprint
  {http://arxiv.org/abs/astro-ph/0303440} {arXiv:astro-ph/0303440 [astro-ph]}
  \BibitemShut {NoStop}%
\bibitem [{\citenamefont {Matos}\ and\ \citenamefont
  {Urena-Lopez}(2007)}]{Matos:2007zza}%
  \BibitemOpen
  \bibfield  {author} {\bibinfo {author} {\bibfnamefont {T.}~\bibnamefont
  {Matos}}\ and\ \bibinfo {author} {\bibfnamefont {L.}~\bibnamefont
  {Urena-Lopez}},\ }\Doi {10.1007/s10714-007-0470-y} {\bibfield  {journal}
  {\bibinfo  {journal} {Gen.Rel.Grav.},\ }\textbf {\bibinfo {volume} {39}},\
  \bibinfo {pages} {1279} (\bibinfo {year} {2007})}\BibitemShut {NoStop}%
\bibitem [{\citenamefont {Bernal}\ \emph {et~al.}(2010)\citenamefont {Bernal},
  \citenamefont {Barranco}, \citenamefont {Alic},\ and\ \citenamefont
  {Palenzuela}}]{Bernal:2009zy}%
  \BibitemOpen
  \bibfield  {author} {\bibinfo {author} {\bibfnamefont {A.}~\bibnamefont
  {Bernal}}, \bibinfo {author} {\bibfnamefont {J.}~\bibnamefont {Barranco}},
  \bibinfo {author} {\bibfnamefont {D.}~\bibnamefont {Alic}}, \ and\ \bibinfo
  {author} {\bibfnamefont {C.}~\bibnamefont {Palenzuela}},\ }\Doi
  {10.1103/PhysRevD.81.044031} {\bibfield  {journal} {\bibinfo  {journal}
  {Phys.Rev.},\ }\textbf {\bibinfo {volume} {D81}},\ \bibinfo {pages} {044031}
  (\bibinfo {year} {2010})},\ \Eprint {http://arxiv.org/abs/0908.2435}
  {arXiv:0908.2435 [gr-qc]} \BibitemShut {NoStop}%
\bibitem [{\citenamefont {Barranco}\ and\ \citenamefont
  {Bernal}(2011)}]{Barranco:2010ib}%
  \BibitemOpen
  \bibfield  {author} {\bibinfo {author} {\bibfnamefont {J.}~\bibnamefont
  {Barranco}}\ and\ \bibinfo {author} {\bibfnamefont {A.}~\bibnamefont
  {Bernal}},\ }\Doi {10.1103/PhysRevD.83.043525} {\bibfield  {journal}
  {\bibinfo  {journal} {Phys.Rev.},\ }\textbf {\bibinfo {volume} {D83}},\
  \bibinfo {pages} {043525} (\bibinfo {year} {2011})},\ \Eprint
  {http://arxiv.org/abs/1001.1769} {arXiv:1001.1769 [astro-ph.CO]} \BibitemShut
  {NoStop}%
\bibitem [{\citenamefont {Urena-Lopez}\ and\ \citenamefont
  {Bernal}(2010)}]{UrenaLopez:2010ur}%
  \BibitemOpen
  \bibfield  {author} {\bibinfo {author} {\bibfnamefont {L.}~\bibnamefont
  {Urena-Lopez}}\ and\ \bibinfo {author} {\bibfnamefont {A.}~\bibnamefont
  {Bernal}},\ }\Doi {10.1103/PhysRevD.82.123535} {\bibfield  {journal}
  {\bibinfo  {journal} {Phys.Rev.},\ }\textbf {\bibinfo {volume} {D82}},\
  \bibinfo {pages} {123535} (\bibinfo {year} {2010})},\ \Eprint
  {http://arxiv.org/abs/1008.1231} {arXiv:1008.1231 [gr-qc]} \BibitemShut
  {NoStop}%
\bibitem [{\citenamefont {Urena-Lopez}\ and\ \citenamefont
  {Liddle}(2002)}]{UrenaLopez:2002du}%
  \BibitemOpen
  \bibfield  {author} {\bibinfo {author} {\bibfnamefont {L.~A.}\ \bibnamefont
  {Urena-Lopez}}\ and\ \bibinfo {author} {\bibfnamefont {A.~R.}\ \bibnamefont
  {Liddle}},\ }\Doi {10.1103/PhysRevD.66.083005} {\bibfield  {journal}
  {\bibinfo  {journal} {Phys. Rev.},\ }\textbf {\bibinfo {volume} {D66}},\
  \bibinfo {pages} {083005} (\bibinfo {year} {2002})},\ \Eprint
  {http://arxiv.org/abs/astro-ph/0207493} {arXiv:astro-ph/0207493} \BibitemShut
  {NoStop}%
\bibitem [{\citenamefont {Cruz-Osorio}\ \emph {et~al.}(2011)\citenamefont
  {Cruz-Osorio}, \citenamefont {Guzman},\ and\ \citenamefont
  {Lora-Clavijo}}]{CruzOsorio:2010qs}%
  \BibitemOpen
  \bibfield  {author} {\bibinfo {author} {\bibfnamefont {A.}~\bibnamefont
  {Cruz-Osorio}}, \bibinfo {author} {\bibfnamefont {F.}~\bibnamefont {Guzman}},
  \ and\ \bibinfo {author} {\bibfnamefont {F.}~\bibnamefont {Lora-Clavijo}},\
  }\Doi {10.1088/1475-7516/2011/06/029} {\bibfield  {journal} {\bibinfo
  {journal} {JCAPA,1106,029.2011},\ }\textbf {\bibinfo {volume} {1106}},\
  \bibinfo {pages} {029} (\bibinfo {year} {2011})},\ \bibinfo {note} {*
  Temporary entry *},\ \Eprint {http://arxiv.org/abs/1008.0027}
  {arXiv:1008.0027 [astro-ph.CO]} \BibitemShut {NoStop}%
\bibitem [{\citenamefont {Marsa}\ and\ \citenamefont
  {Choptuik}(1996)}]{Marsa:1996fa}%
  \BibitemOpen
  \bibfield  {author} {\bibinfo {author} {\bibfnamefont {R.~L.}\ \bibnamefont
  {Marsa}}\ and\ \bibinfo {author} {\bibfnamefont {M.~W.}\ \bibnamefont
  {Choptuik}},\ }\href@noop {} {\bibfield  {journal} {\bibinfo  {journal}
  {Phys. Rev.},\ }\textbf {\bibinfo {volume} {D54}},\ \bibinfo {pages} {4929}
  (\bibinfo {year} {1996})},\ \Eprint {http://arxiv.org/abs/gr-qc/9607034}
  {gr-qc/9607034} \BibitemShut {NoStop}%
\bibitem [{\citenamefont {Thornburg}(1999)}]{Thornburg:1998cx}%
  \BibitemOpen
  \bibfield  {author} {\bibinfo {author} {\bibfnamefont {J.}~\bibnamefont
  {Thornburg}},\ }\Doi {10.1103/PhysRevD.59.104007} {\bibfield  {journal}
  {\bibinfo  {journal} {Phys. Rev.},\ }\textbf {\bibinfo {volume} {D59}},\
  \bibinfo {pages} {104007} (\bibinfo {year} {1999})},\ \Eprint
  {http://arxiv.org/abs/gr-qc/9801087} {arXiv:gr-qc/9801087} \BibitemShut
  {NoStop}%
\bibitem [{Cho(2003)}]{Choptuik:2003}%
  \BibitemOpen
  \href@noop {} {\enquote {\bibinfo {title} {{Graduate Summer School on General
  Relativistic Hydrodynamics}},}\ } (\bibinfo {year} {2003}),\ \bibinfo {note}
  {http://cgwp.gravity.psu.edu/events/GRHydro03/}\BibitemShut {NoStop}%
\bibitem [{\citenamefont {Alcubierre}(2008)}]{Alcubierre08a}%
  \BibitemOpen
  \bibfield  {author} {\bibinfo {author} {\bibfnamefont {M.}~\bibnamefont
  {Alcubierre}},\ }\href@noop {} {\emph {\bibinfo {title} {{Introduction to
  $3+1$ Numerical Relativity}}}}\ (\bibinfo  {publisher} {Oxford Univ. Press,
  New York},\ \bibinfo {year} {2008})\BibitemShut {NoStop}%
\bibitem [{\citenamefont {Hernandez}\ \emph {et~al.}(2009)\citenamefont
  {Hernandez}, \citenamefont {Mendoza}, \citenamefont {Rendon}, \citenamefont
  {Lopez-Monsalvo},\ and\ \citenamefont {Velasco-Segura}}]{Hernandez:2007jt}%
  \BibitemOpen
  \bibfield  {author} {\bibinfo {author} {\bibfnamefont {X.}~\bibnamefont
  {Hernandez}}, \bibinfo {author} {\bibfnamefont {S.}~\bibnamefont {Mendoza}},
  \bibinfo {author} {\bibfnamefont {P.~L.}\ \bibnamefont {Rendon}}, \bibinfo
  {author} {\bibfnamefont {C.~S.}\ \bibnamefont {Lopez-Monsalvo}}, \ and\
  \bibinfo {author} {\bibfnamefont {R.}~\bibnamefont {Velasco-Segura}},\ }\Doi
  {10.3390/e11010017} {\bibfield  {journal} {\bibinfo  {journal} {Entropy},\
  }\textbf {\bibinfo {volume} {11}},\ \bibinfo {pages} {17} (\bibinfo {year}
  {2009})},\ \Eprint {http://arxiv.org/abs/gr-qc/0701165} {arXiv:gr-qc/0701165}
  \BibitemShut {NoStop}%
\end{thebibliography}%

\end{document}